# Surface Properties of Ga-Cu Based Liquid-Metal Alloys: Impact of Cu Dilution, Topography, and Reaction Conditions

Tzung-En Hsieh[1,†,*], Michael S. Moritz[2,†], Andreas Mölkner[3,†], Christoph Wichmann[2,4], Johannes Frisch[1,5], Julien Steffen[3], Caiden J. Parker[6], Vaishnavi Krishnamurthi[6], Torben Daeneke[6], Hans-Peter Steinrück[2], Andreas Görling[3,7], Christian Papp[4], Marcus Bär[1,5,8,9,*]

[1] Department Interface Design, Helmholtz-Zentrum Berlin für Materialien und Energie GmbH (HZB), 12489 Berlin, Germany

[2] Friedrich-Alexander-Universität Erlangen-Nürnberg (FAU), Lehrstuhl für Physikalische Chemie, 91058 Erlangen, Germany

[3] Friedrich-Alexander-Universität Erlangen-Nürnberg (FAU), Lehrstuhl für Theoretische Chemie, 91058 Erlangen, Germany.

[4] Freie Universität Berlin, Angewandte Physikalische Chemie, 14195 Berlin, Germany

[5] Energy Materials In-situ Laboratory Berlin (EMIL), HZB, 12489 Berlin, Germany

[6] School of Engineering, RMIT University, 3000 Melbourne, Australia

[7] Erlangen National High Performance Computing Center (NHR@FAU), D-91058 Erlangen, Germany

[8] Department of Chemistry and Pharmacy, Friedrich-Alexander-Universität Erlangen-Nürnberg (FAU), 91058 Erlangen, Germany

[9] Department X-ray Spectroscopy at Interfaces of Thin Films, Helmholtz-Institute Erlangen-Nürnberg for Renewable Energy (HI ERN), 12489 Berlin, Germany

[†]The authors contributed equally to this work.

*Corresponding authors: Tzung-En Hsieh: tzung-en.hsieh@helmholtz-berlin.de, Marcus Bär: marcus.baer@helmholtz-berlin.de

**Table of contents:**

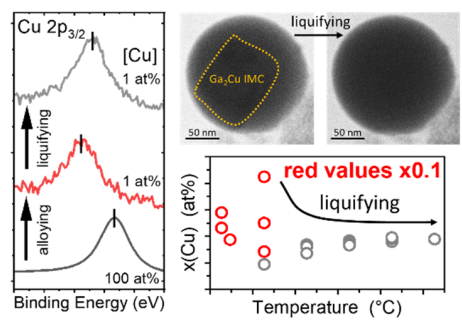

# Abstract

We studied the surface properties of Ga-Cu based liquid metal alloys – a promising material system for supported catalytically active liquid metal solutions (SCALMS). The impact of Cu dilution in the (liquid) Ga matrix is in-detail investigated by X-ray and UV photoelectron spectroscopy (XPS/UPS) and Machine-Learned-Force Field (ML-FF) calculations. With decreasing Cu content, microscopic and macroscopic Ga-Cu model samples exhibit a shift of the Cu 2p core level line to higher binding energies ($E_b$) as well as a correspondingly shifted and narrowed d-band with respect to pure Cu, which we ascribe to site isolation. To study the property evolution of Ga-Cu at SCALMS reaction conditions, i.e., where Cu is present in liquid Ga, additional XPS measurements were performed between 100 and 500 °C. The observed Cu 2p shift to lower $E_b$ is tentatively ascribed to changes in the local environment with increasing temperature, i.e. bond elongation, which is corroborated by ML-FF simulations; the increased Cu surface content at low temperatures is attributed to the presence of crystallized Cu-rich intermetallic compounds, as evidenced by transmission electron microscopy images. In an attempt to generalize the findings for filled d-band transition metals (TMs) in liquid Ga also Ga-Ag and Ga-Au model systems were investigated. We also find indications for site isolation, however, with a different segregation behavior of the dissolved TM atoms. The observed insights may be another step of paving the way for an insight-driven development of low-temperature melting liquid metals for heterogeneous catalysis.

# Introduction

Research in catalysis generally aims to improve industrially relevant reactions to be more economic, resource-efficient, and productive. Catalysis is divided into two categories, homogeneous and heterogeneous catalysis.[1] In homogeneous catalysis, the catalyst often consists of a transition metal complex that is soluble in the same solution as the reactant molecules, while the reactants and the catalyst are in separate phases in heterogeneous catalysis. In the latter, the reaction occurs at the interface of catalysts and reactants.[2] Heterogeneous catalysts have a high potential for industrial applications and are utilized more frequently because of their high scalability and efficient catalyst-product separation.[1] However, a challenge of heterogeneous catalysts is the accurate characterization of the active sites under reaction conditions. Often, the surface is composed of complicated configurations of active sites and intermediates that only exist at operating conditions. Therefore, models that translate the scarce information obtained on surface reactions (by analytical methods that are often not compatible with relevant conditions) have limited relevance, especially when proposing possible reaction mechanisms.[3-4] A general concept that, however, is quite successfully used to explain the structure-function relationship of catalysts, is the d-band model by Hammer and Nørskov.[5-7] According to this model, the width and the position of the d-band with respect to the Fermi-level and the frontier orbital of adsorbates, i.e., reactants, intermediates, and products, are associated with the binding strength and hence are related to the catalytic activity and selectivity. A deliberately optimized active species with an ideal d-band thus induces a modified binding strength with adsorbates, leading to an improvement in catalytic performance. This model has revolutionized modern catalysis research and explains many empirically derived trends.[6, 8-9]

According to previous studies, a typical strategy to tailor the d-band of transition metals is alloying.[10] Upon alloying, the chemical environment of transition metals is modified from delocalized metal bonding into the localized intermetallic bonding with pronounced covalent characteristics. This modification of chemical environments drastically alters the chemical and electronic structure of transition metal atoms, inducing a modification of the d-band. Recently, one novel approach to create these binary species has been introduced: In Supported Catalytically Active Liquid Metal Solutions (SCALMS), the catalytically active transition metals, e.g., Pt, Rh, Pd, Ni, are dispersed in a low-melting-point liquid metal matrix (e.g., Ga) in low concentrations (<5 at.%).[11-14] This leads to the formation of highly sustainable catalysts with isolated transition metal atoms as active sites.[15-16] Recent studies showed the coking- and sintering-resistant liquid metal nature of the catalyst surfaces under reaction conditions.[17-18] In addition, in previous work, we reported on a temperature-dependent formation and dissolution of multiple binary species in Ga-Rh and Ga-Ni SCALMS model systems, which leads to a variation of the catalytic performance upon changing the pre-treatment and reaction conditions. This indicates that in some cases more than one potential active species can be present during the reactions.[19-20] Also, it has been reported that the liquid metal matrix may affect the fundamental properties of the active sites in SCALMS.[21]

In order to complete the study on Ga-based binary SCALMS systems, in this paper, we interrogate the chemical and electronic structure of filled d-band transition metals in a Ga matrix. Ga-Cu and Ga-Ag are utilized as binary catalysts and are considered as potential constituents in ternary SCALMS.[22-23] Hence, we have investigated binary Ga-Cu, Ga-Ag, and Ga-Au model systems, with a clear focus (because of the expected wide-spread application) on the Ga-Cu system. We have in-detail studied the impact of Cu dilution and liquefaction on the chemical and electronic structure of Ga-Cu samples by photoemission measurements and transmission electron microscopy (TEM) in combination with machine-learned-force field (ML-FF) and density function theory (DFT) calculations. The insights on how the filled Cu d-band determines the chemical structure have been complemented by results derived from exemplary photoemission measurements on binary Ga-Ag and Ga-Au samples. In order to further study the influence of the topography on the catalyst properties, we have investigated microscopic and macroscopic model systems (see Fig. S1). Ultimately, the gained combined insights are expected to pave the way for deliberately developing novel liquid metal catalysts.

## Results and Discussions

### Impact of Cu dilution on the chemical and electronic properties

Recently, a narrowing of the Cu 3d-derived band was determined to greatly influence the binding strength to methanol, altering activation energies of methanol reforming for single-atom-like Cu in Ag.[24] For SCALMS, it was similarly shown that there is a relation between the electronic structure of the active species and the catalytic productivity.[20-21] Herein, a series of microscopic Ga-Cu model systems with different Cu contents were prepared via Physical-Vapor-Deposition (PVD). To elucidate the impact of Cu dilution on the electronic and chemical structure of Cu in a Ga matrix at room temperature (r.t.), the samples were investigated by X-ray and ultraviolet photoelectron spectroscopy (XPS and UPS). Survey spectra of the clean samples are shown in Fig. S2. The metallic nature of the Ga matrix surface is confirmed by fit analysis of the Ga 3d and O 1s core level spectra of the Ga-Cu samples (Fig. S2c, S3). The oxygen signal shown in Fig. S2c is attributed to the exposed $SiO_x$/Si support.[25] Fig. 1a shows the Cu $2p_{3/2}$ core level peak of pure Cu compared to that of microscopic Ga-Cu alloy samples with 36 at%, 5 at%, and 1 at% Cu. The pure Cu reference shows a single-species Cu $2p_{3/2}$ line at 932.6 eV. Upon decreasing the Cu content in the Ga matrix to 36 at%, a Cu $2p_{3/2}$ core level shift of +1.0 eV (933.6 eV) to higher binding energy

($E_b$) is observed. This Cu $2p_{3/2}$ peak position and shape are preserved upon further Cu content decrease to 5 and 1 at% (see Fig. S4), indicating the formation of a stable Ga-Cu chemical environment that is different from that of pure metallic Cu. In addition, we rule out the formation of Cu-O on the sample surface based on the observed oxide-free Ga surface shown in Fig. S2c and Fig. S3. This step-wise change is different from our Ga-Rh,[19, 26] Ga-Pt,[27] and Ga-Ni[20] studies where we observed a continuous (sometimes complex) shift of the XPS TM core levels upon TM dilution.

The UPS spectra reveal the evolution of the valence electronic structure of the microscopic Ga-Cu model systems upon Cu dilution in Ga (Fig. 1b). The PVD-prepared pure Cu sample shows a Cu 3d-derived spectral feature centered at 3 eV with an approximate FWHM of 2 eV. Upon decreasing the Cu concentration to 36 at%, the Cu 3d band center shifts by +1.0 eV to higher $E_b$ (in agreement with the observed Cu 2p shift), and the FWHM narrows by 0.5 eV. A further decrease in the Cu content to 5 and 1 at%, leads to a similar Cu 3d-derived spectral feature at 3.5-5 eV (see detailed evaluation of corresponding difference spectra in Fig. S5). The higher spectral intensity at 5.5 eV in the spectrum of the 1 at% Cu sample is tentatively attributed to O 2p derived states caused by oxidized Cu (due to interaction with the support) and/or by the exposed $SiO_x$/Si support itself.

In order to understand the changes of the spectral d-band-derived feature in the valence band (VB) due to Cu dilution, ML-FF calculations and DFT calculations of the unit cells with 180 atoms are employed (see details in SI). The calculated Cu PDOS of a Cu (111) single crystal shows a broad feature between 2-5 eV. In the initial structure of the ML-FF calculations for the Ga-Cu model samples containing 36 at% and 1 at% Cu, the atoms in the unit cells are randomly set (random alloy model). During the simulations, no formation of a crystal structure was observed. Upon decreasing the Cu content to 36 at% and 1 at%, the Cu PDOS narrows with respect to that of Cu (111) and is centered at 3 eV. We attribute the changes of Cu PDOS to the new chemical environment of diminishing Cu-Cu interaction in the calculated unit cell. According to the snapshot shown in Fig. S6, no Cu-Cu interaction is observed in the unit cell containing 1 at% Cu, showing that the Cu atoms are always exclusively surrounded by Ga atoms, independent of their position in the cell and temperature. Note that the Cu PDOS for the unit cell with 36 at% Cu shows a broader PDOS than the 1 at% Cu model, which is attributed to a higher possibility of Cu-Cu interactions in the unit cell (Fig. S6).

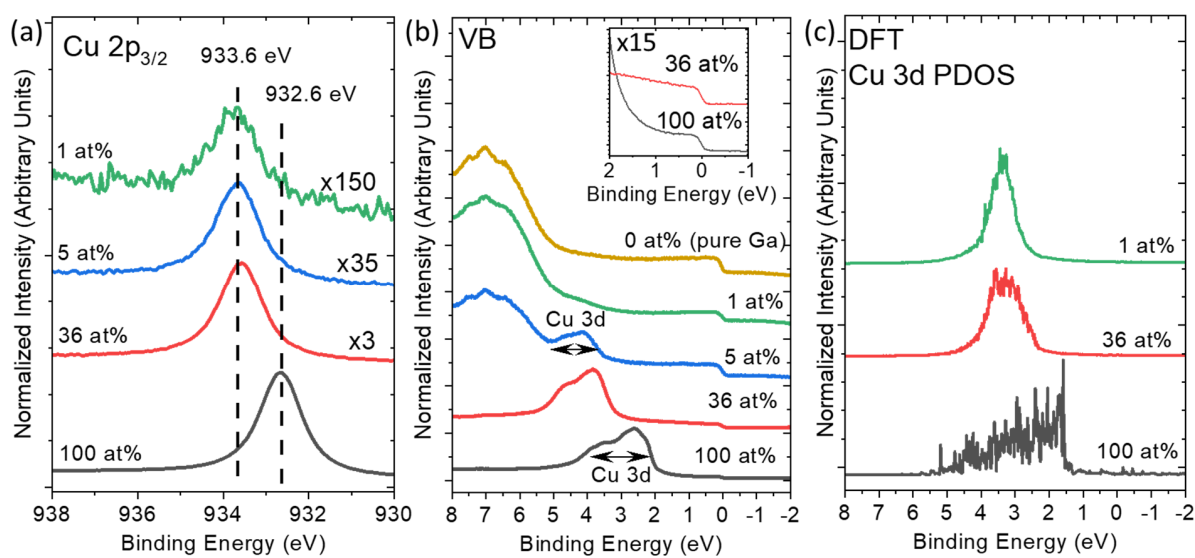

**Figure 1.** (a) Mg K$_\alpha$-excited XPS Cu 2p$_{3/2}$ detail spectra and (b) He II excited UPS VB spectra of PVD-prepared SiO$_x$/Si-supported Ga-Cu microscopic model system samples with Cu contents of 100, 36, 5, and 1 at% (from bottom to top spectrum) measured at r.t.. Note the different magnification factors. The Cu 2p$_{3/2}$ E$_b$ of metallic Cu (932.6±0.1 eV)[28] and of the newly formed Ga-Cu species (933.6±0.1 eV)[29] are indicated for reference. Panel (b) shows the UPS spectra of the VB region of the samples shown in (a) in comparison to that of pure Ga (0 at%). The Fermi edge at E$_b$ = 0 eV of the 100 and 36 at% containing Cu sample is shown in the inset on a magnified scale of ×15. (c) DFT-calculated Cu 3d-derived partial density of states (PDOS) of a Cu (111) surface and Ga-Cu random alloys containing 36, and 1 at% Cu at room temperature. The PDOS feature in (c) is normalized by the total feature area. In all panels the spectra are depicted with a constant offset for better visibility.

According to previous studies, another explanation for the evolution of the spectral UPS and XPS features upon Cu dilution is the attribution of the new Cu environment as being similar to that of a $Ga_2Cu$ intermetallic compound (IMC).[23] For the microscopic Ga-Cu sample with the highest Cu content, this is in line with the Cu concentration of 36 at%, which is close to the nominal Cu concentration of 33 at% in $Ga_2Cu$ and a calculated Cu $2p_{3/2}$ core level $E_b$ position of $933.5\pm0.1$ eV at r.t. (Fig. S7). However, due to the low Cu content in the 5 at% and 1 at% Cu containing model systems, the similar Cu $2p_{3/2}$ $E_b$ observed in these samples might have origins other than $Ga_2Cu$ IMC formation. As discussed above, ML-FF simulations and DFT calculations demonstrate that the Cu dilution in the random set Ga-Cu unit cell can lead to a shift and narrowing of the Cu 3d derived PDOS, which is in agreement with the UPS results (see Fig. 1b). This behavior is attributed to a decrease in the number of Cu-Cu bonds. Similar (core level) shifts to higher $E_b$ and narrowing d-bands were also observed upon Rh and Ni dilution in Ga-Rh and Ga-Ni samples, for which spatial site isolation (i.e., the absence of bonds between TM atoms) has been suggested as an explanation.[20, 26] According to the crystal structure of $Ga_2Cu$, also in this case Cu is forming chemical bonds only with Ga.[30] Thus, both the formation of $Ga_2Cu$ IMC and/or the presence of a material best described by the random alloy model calculated by ML-FF are in agreement with the shift of the Cu 2p and Cu derived d-band (and narrowing of the latter) caused by a significant decrease of the number of Cu-Cu bonds (which one could interpret as site isolation of the TM). In consequence, we speculate that in the PVD-grown Ga-Cu microscopic model samples, $Ga_2Cu$ IMC-like regions may coexist with regions best described by the calculated random alloy model.

**Evolution of the Cu chemical environment upon liquefaction**

To probe the chemical structure of Cu in liquid Ga, we elucidate the impact of annealing on the Ga-Cu microscopic model system via temperature-dependent XPS measurements. The 1 at% Cu containing Ga-Cu model system is measured with XPS at r.t., 200, and 400 °C (Fig. 2a-c). Fig. 2a shows the Cu $2p_{3/2}$ region measured at r.t. featuring a dominant species at 933.6 eV. Fit analysis – based on the model established in conjunction with Fig. S4 – reveals the presence of two components. The minor contribution (fit component Cu_1), we ascribe to metallic Cu and the main contribution (Cu_2) is attributed to site-isolated Cu in the form of a $Ga_2Cu$ IMC and/or a material best described by the ML-FF derived random alloy model. A consecutive Cu $2p_{3/2}$ core level shift upon annealing is observed. At 200 °C, i.e., a temperature lower than the melting point of $Ga_2Cu$ (derived to be 261.26°C by differential scanning calorimetry measurements (DSC) in agreement with Refs.[31-32] – see Fig. S8), the major fit component (Cu_2) shifts to lower $E_b$ by -0.1 eV. An additional shift of -0.2 eV to lower $E_b$ is observed upon further temperature increase to 400 °C. The observed spectral variations upon annealing could be related to the slight alternation and complete dissolution of the $Ga_2Cu$ IMC at 168 °C and 268 °C, respectively. (see Fig. S9 and Table S1). This IMC dissolution presumably alters Cu-Ga bond lengths and angles in the sample in agreement with the broadening of the calculated spectra upon temperature increase. Note, the contribution of metallic Cu is not increasing upon annealing procedures (< 5% of the total intensity).

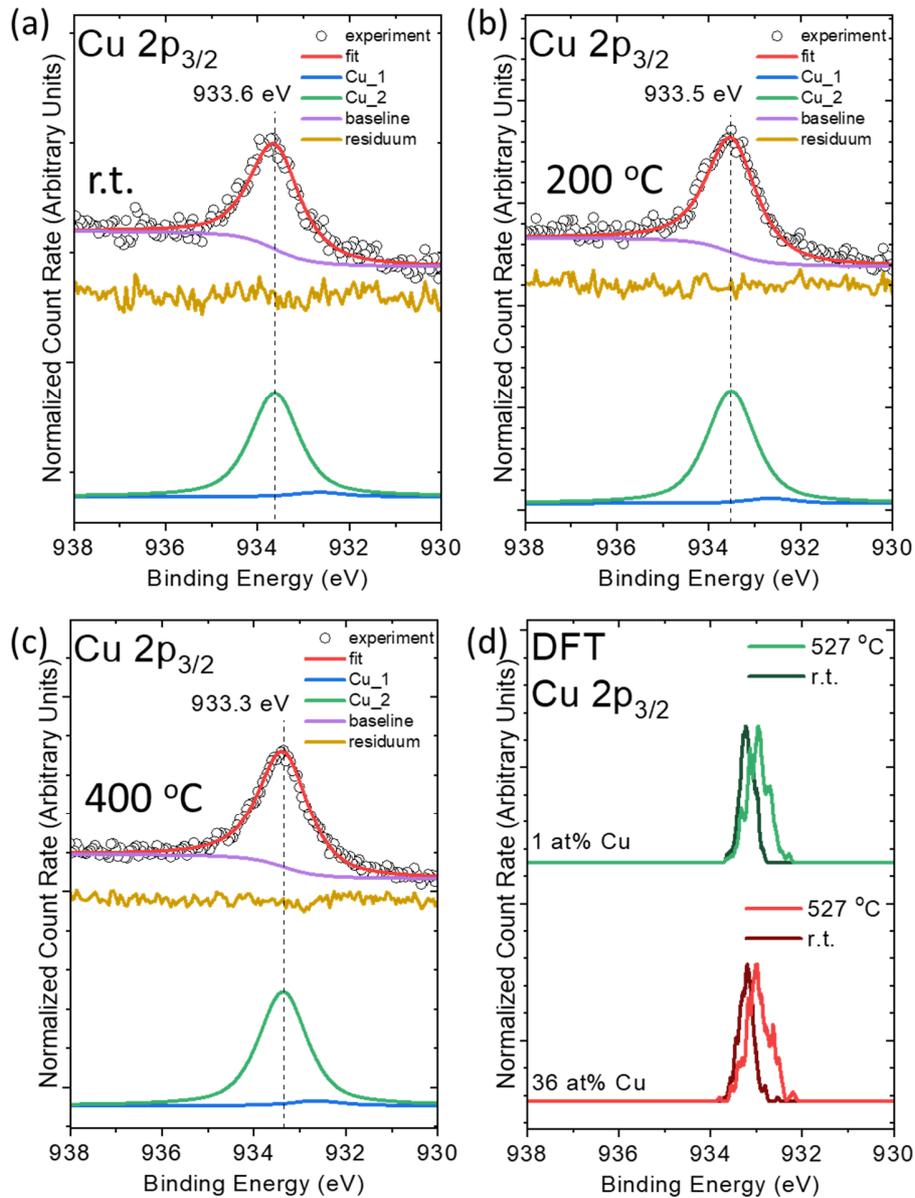

**Figure 2.** Fitting results of Mg K$_\alpha$-excited XPS Cu 2p$_{3/2}$ spectra of PVD-prepared SiO$_x$/Si-supported Ga-Cu microscopic model system sample with 1 at% Cu measured in-situ at r.t. (a), 200 °C (b), and 400 °C (c), respectively. (d) DFT-calculated Cu 2p$_{3/2}$ core level peak position of Cu atoms in a Ga-Cu random alloy model with 36 at% and 1 at% Cu contents at r.t. and 527 °C.

Fig. 2d shows Cu 2p$_{3/2}$ core level positions calculated by DFT for Ga-Cu unit cells with 36 at% and 1 at% Cu randomly distributed in a Ga matrix (random alloy model) at r.t. (27 °C) and 527 °C. Upon annealing, a core level peak broadening and shift to lower E$_b$ with respect to r.t. is observed which corroborates the experimental XPS results (Fig. 2a-c). Similarly, a Cu $2p_{3/2}$ shift to lower $E_b$ and broadening is also calculated for $Ga_2Cu$ IMC upon annealing to 527 °C (Fig. S7). According to the calculated radial distribution functions (RDF) shown in Figs. S10-11 and S12 for the random alloy model and the $Ga_2Cu$ IMC, respectively, the average Ga-Cu distance increases upon annealing in both 1 at% and 36 at% Cu containing Ga-Cu unit cells (see SI). The shift and broadening of the Cu $2p_{3/2}$ level in the calculation might hence be attributed to changes in the Ga-Cu bond length. The impact of the lattice expansion altering the bond length and angle has been reported in previous studies.[24, 33] In literature, the dynamic dissolution of $Ga_2Cu$ in the Ga matrix was observed by in-situ TEM and EDS measurements during annealing,[34] showing a dispersion of Cu atoms in the entire Ga droplet. In contrast to the Cu 2p XPS line showing obvious changes (Fig. 2a-c), the Ga 3d core level peak shape and position are preserved independent of temperature, showing no indication of Ga oxidation upon annealing (Fig. S13).

Since the Ga-Cu microscopic model system samples suffer from de-wetting (see Fig. S14), a macroscopic Ga-Cu droplet was prepared and studied in order to minimize the effects induced by sample topography. XPS measurements were conducted on a macroscopic (cm size, see Fig. S1) Ga-Cu droplet with a nominal Cu content of 1 at% (Fig. 3). Fig. 3a shows the Ga 3d spectra obtained at 27°C, which displays a single asymmetric peak that is fitted by a Ga $3d_{5/2}$ component at 18.4 eV and a corresponding spin-orbit-split Ga $3d_{3/2}$ component at 0.46 eV higher $E_b$. No spectral change is observed at 427°C (Fig. 3b). The samples are gallium oxide-free, as no $Ga^{3+}$ contributions at 20.4 eV are observed (Fig. 3a, b, Fig. S15, S16b), in agreement with the results on the microscopic Ga-Cu model system discussed above. Fig. 3c shows a single species at 933.7 eV in the Cu $2p_{3/2}$ spectra at 27 °C (note, the intensity was lowered by a factor of ×0.1 to allow for easy comparison; high Cu-intensities were measured due to IMC surface crystallization, as discussed later). At 427 °C, the Cu $2p_{3/2}$ peak position is shifted to 933.3 eV by -0.4 eV to lower $E_b$ (Fig. 3d), both the Cu $2p_{3/2}$ $E_b$ position and the temperature-induced shift corroborate the XPS derived findings of the microscopic model system (see Fig. 2a, c) and the DFT calculations. Fig. 3e summarizes the evolution of the Ga $3d_{5/2}$ and Cu $2p_{3/2}$ $E_b$ of both microscopic and macroscopic model system samples with a nominal Cu content of 1 at% during temperature ramping from 27 °C to 527 °C (raw data shown in Fig. S16). Similar chemical structure changes indicate a negligible influence of the support or sample topography.

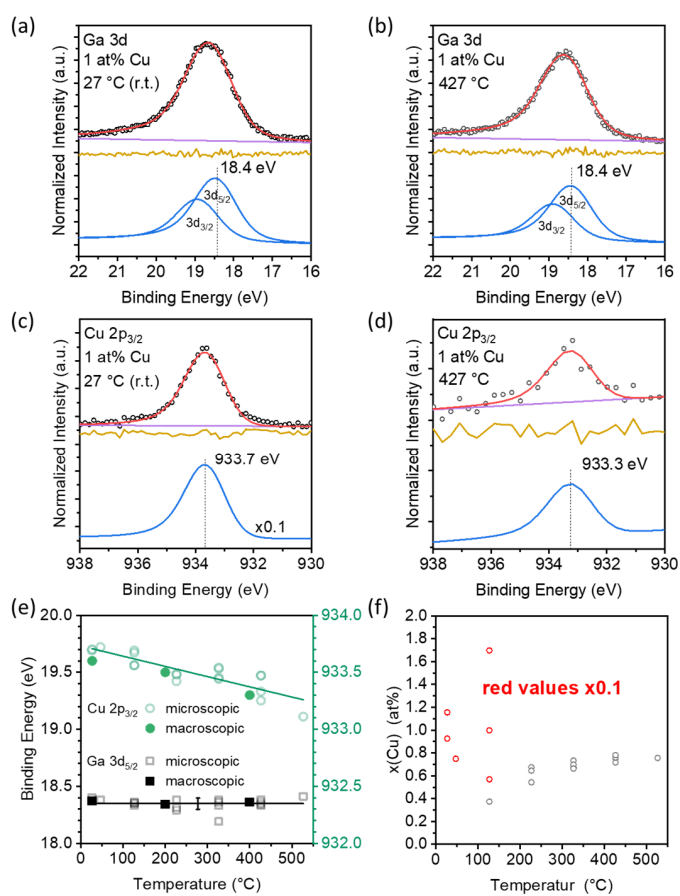

**Figure 3.** Al $K_\alpha$-excited (a, b) Ga 3d and (c, d) Cu $2p_{3/2}$ XPS details spectra of a macroscopic Ga-Cu model system sample with a nominal Cu content of 1 at% measured at 27 and 427 °C. Note the magnification scale of ×0.1 for the r.t. Cu $2p_{3/2}$ spectrum shown in (c). (e) Evolution of the Ga $3d_{5/2}$ and Cu $2p_{3/2}$ core level peak position for microscopic (closed symbols) and macroscopic (open symbols) Ga-Cu model system samples with a nominal Cu content of 1 at% as a function of temperature. (f) Evolution of the Cu surface

content of a macroscopic model system sample with a nominal Cu content of 1 at% as a function of temperature.

The surface phase behavior of alloys can be significantly different from the bulk behavior because of segregation effects to or away from the gas/liquid interface.[35] Thus, investigating the surface-derived phase diagram of the alloys is crucial for understanding the behavior of surface atoms, i.e., recrystallization, dealloying, etc., upon annealing.[36] In Fig. 3f, the temperature induced evolution of the XPS derived (see details in SI) surface Cu content is depicted. At 27 °C, the calculated Cu concentration is between 7 and 12 at% (9.4 ± 2.0 at%), which is significantly higher than the nominal 1 at% bulk concentration derived from weigh-in (see experimental section). This Cu enrichment can be explained by the crystallization of Cu-rich Ga-Cu IMCs, e.g. $Ga_2Cu$, at the droplet surface,[37] as corroborated by the corresponding TEM image in Fig. S9. The presence of $Ga_2Cu$ in the Ga matrix at room temperature has also been previously observed in TEM[34] and discussed as a possible origin of the observed Cu $2p_{3/2}$ shift to higher $E_b$ upon Cu dilution in the Ga-Cu microscopic model system above. At 127 °C, the surface Cu concentration varies between 15 and 0.3 at% (8.2 ± 6.9 at%) for repeated experiments shown in Fig. 3f. This large variation could be explained by an increased mobility of the Ga-Cu IMCs (allowing for the IMC moving out of the XPS probing area). At 227 °C and above, the concentration of Cu in the surface-near region is 0.7 ± 0.1 at%. Our DSC measurements indicate a melting point of roughly 261.26 °C for $Ga_2Cu$ (see Fig. S8) and TEM analysis suggests that dissolution already starts at temperatures as low as 170 °C (see Fig. S9 and Table S1). We therefore attribute this plateau to the full liquefaction of the system. Compared to the bulk concentration of 1 at%, the surface is Cu-depleted in this temperature regime. A similar behavior was observed for Ga-Pt alloys.[21] Interestingly, the Cu surface concentration is not correlated with the Cu $2p_{3/2}$ core level shift seen in Fig. 3e. This indicates that the shift could be correlated with the changes in the local environment with increasing temperature, i.e. bond elongation, predicted by the ML-FF simulations and not solely with the melting of an intermetallic phase (Fig. S10-12). In the latter case, a step-wise change at a distinct temperature would be expected, while the found linear shift is best explained by temperature-dependent alterations of Cu-Cu and Ga-Cu bonds as discussed in conjunction with Fig. S6, Fig. S12b and previously for the microscopic model system.

Besides the in detail interrogated Ga-Cu system, we have also elucidated the properties of Ga-Ag- and Ga-Au-based microscopic and macroscopic model system samples, i.e. Ga matrices with other filled d-band elements (See SI, Fig. S17-27). First results for Ga-Ag and Ga-Au samples with 1 at% Ag and 1 at% Au, respectively, show similar core level shifts for the transition metals toward higher $E_b$ upon dilution, which can be tentatively interpreted as indication for site isolation (Fig. S17), as discussed in-detail for the Ga-Cu model system. Nevertheless, the surface concentration of Ag and Au varies differently upon alloy liquefaction as compared to Cu. In contrast to the Ga-Cu system, the surface Ag concentration shows negligible changes during annealing (Fig. S23) while the Au content increases when the annealing temperature reaches the theoretical melting point.[31] These observed differences of the three model systems based on filled d-band transition metals indicate that more detailed studies and discussions have to be conducted for a complete fundamental picture of these liquid metal alloys, as a prerequisite for a deliberate development as liquid metal catalysts.

## Conclusion

By investigating Ga-Cu microscopic and macroscopic model systems by XPS and UPS, this study sheds light on the fundamental properties of potential liquid metal alloy-based catalysts. The observed spectral line changes including the Cu $2p_{3/2}$ XPS peak shift and the UPS-derived d-band narrowing with respect to a pure Cu reference upon Cu dilution, are attributed to Ga-Cu alloying already at room temperature. In view of previous studies and the presented ML-FF calculations, these spectral changes suggest the formation of site-isolated Cu atoms, in either the form of a $Ga_2Cu$-type IMC and/or a material best described by the calculated random alloy phase. Furthermore, temperature-dependent XPS experiments elucidated the variation of the chemical structure of the Cu atoms in microscopic and macroscopic model systems upon sample liquefaction, i.e. at reaction-relevant temperatures at which the sample is expected to be in its liquid phase. A temperature-dependent Cu $2p_{3/2}$ core level peak shift with increasing temperature is observed, which is interpreted as an indication for the dissolution of $Ga_2Cu$ in the liquid Ga matrix (as corroborated by TEM) and the alteration of bond lengths upon annealing. Measurements of additional Ga – filled d-band TM (i.e., Ga-Ag and Ga-Au) model systems reveal similar core level shifts to higher $E_b$ upon Ag and Au dilution in Ga (interpreted as an indication for site isolation). However, our temperature-dependent XPS experiments suggest different Ag and Au surface content evolutions upon sample liquefaction, indicating a complex situation. Irrespectively, our studies of the chemical and electronic structure of filled d-band TMs in Ga-based alloys provide fundamental insights crucially required for a knowledge driven development of related SCALMS catalysts.

**AUTHOR INFORMATION**

The authors declare no conflict of interests.

**AUTHOR CONTRIBUTIONS**

‡These authors contributed equally. T.H., M.M., and A.M. designed and realized the experiments and calculations, performed synthesis, characterization, analysis, interpretation, wrote the original draft, reviewed, and edited the final manuscript. J.S supported DFT and ML-FF calculations, interpretation, reviewed and edited the final manuscript. C.J.P conducted TEM investigations. V.K. was responsible for the DSC measurements. T.D. initiated and interpreted the TEM and DSC results together with V.K. and C.J.P. C.W., J.F., and J.S. performed interpretation and reviewed the final manuscript, M.B., C.P., A.G., and H.-P.S. developed the scientific concept, acquired the funding, reviewed, and edited the final manuscript. All authors contributed to the discussion and analysis of the results presented and have given approval to the final version of the manuscript.

**Keywords:** liquid metal alloys, SCALMS, gallium, copper, silver, gold, XPS, UPS, DFT, surface phase diagram, ML-FF

**Supporting Information:**

Description of sample preparation and ML-FF calculations; XPS peak analysis procedure; additional XPS/UPS spectra of Ga-Cu, Ga-Ag and Ga-Au alloys; Ga 3d, Cu $2p_{3/2}$, Ag 3d, and Au 4f XPS fitting results; additional simulations results; additional spectroscopic results about Ga-Ag and Ga-Au model systems.

**Data Access:**

The dataset of XPS spectra present in this manuscript and supporting information is accessible with following Zenodo DOI: 10.5281/zenodo.17019725

**ACKNOWLEDGMENT**

C.P. thank for financial support by the European Research Council through Project 786475: Engineering of Supported Catalytically Active Liquid Metal Solutions. Additional infrastructural and financial support by the German Research Foundation (DFG) through the collaborative research centre CRC 1452 is gratefully acknowledged. The authors also gratefully acknowledge the scientific support and HPC resources provided by the Erlangen National High Performance Computing Center (NHR@FAU) of the Friedrich-Alexander-Universität Erlangen-Nürnberg (FAU) under the NHR project b146dc. NHR funding is provided by federal and Bavarian state authorities. NHR@FAU hardware is partially funded by the German Research Foundation (DFG) – 440719683. T.D, V.K. and C.J.P thank the Australian Research Council for funding via the Discovery Scheme (DP220101923 and DP240101215) and acknowledge the facilities and the technical assistance of RMIT University's Microscopy and Microanalysis Facility (RMMF). The authors thank to Dr. Florian Ruske for his supports on SEM measurements of microscopic model systems. EMIL is also acknowledged for making the infrastructure available for the in-system sample preparation and lab-based XPS and UPS measurements.